# MENTORING SOFTWARE IN EDUCATION AND ITS IMPACT ON TEACHER DEVELOPMENT: AN INTEGRATIVE LITERATURE REVIEW

Ramiro Pesina

The University of Texas at Tyler, USA

***ABSTRACT***

*Mentoring software is a pivotal innovation in addressing critical challenges in teacher development within educational institutions. This study explores the transformative potential of digital mentoring platforms, evaluating their impact on enhancing traditional mentoring practices through scalable, data-driven, and accessible frameworks. The research synthesizes findings from existing literature to assess the effectiveness of key features, including structured goal setting, progress monitoring, and advanced analytics, in improving teacher satisfaction, retention, and professional growth.*

*Using an integrative literature review approach, this study identifies both the advantages and barriers to implementing mentoring software in education. Financial constraints, limited institutional support, and data privacy concerns remain significant challenges, necessitating strategic interventions. Drawing insights from successful applications in healthcare and corporate sectors, the review highlights adaptive strategies such as leveraging open-source tools, cross-sector collaborations, and integrating mentoring software with existing professional development frameworks.*

*The research emphasizes the necessity of integrating digital mentoring tools with institutional objectives to create enduring support systems for teacher development. Mentoring software not only enhances traditional mentorship but also facilitates broader professional networks that contribute to collective knowledge sharing.*

## 1. MENTORING SOFTWARE IN EDUCATION AND ITS IMPACT ON TEACHER DEVELOPMENT: AN INTEGRATIVE LITERATURE REVIEW

Mentoring plays a critical role in education by equipping teachers with the necessary tools and confidence to become highly effective educators. Research indicates that mentoring is particularly essential for new educators, as it provides support in navigating classroom challenges, developing curriculum strategies, and fostering strong connections with the school community (Enciso, 2024; Huang et al., 2024). However, many educational institutions struggle to sustain effective mentoring programs due to financial constraints, limited institutional support, and logistical inefficiencies (Mumo et al., 2024; Zamiri& Esmaeili, 2024). The lack of adequate mentoring resources often contributes to high teacher attrition rates and professional dissatisfaction, further exacerbating the challenges faced by schools in retaining skilled educators (Anis, 2024; Maxwell et al., 2024).

The increasing adoption of digital tools in education has generated interest in how technology can be leveraged to enhance traditional mentorship models. Digital mentoring platforms provide structured, scalable solutions that can address gaps in traditional mentoring programs by utilizing data-driven methodologies, real-time communication tools, and automated progress tracking (Bagai & Mane, 2024; Choudhary et al., 2024). Similar approaches have been successfully implemented in higher education, professional training, and career development, demonstrating potential applications for teacher mentorship (Onyema et al., 2024; Ovais & Jain, 2025). Additionally, peer mentoring programs supported by digital platforms have been shown to enhance professional development, providing ongoing support for educators in training and throughout their careers (Gehreke, Schilling, & Kauffeld, 2024). Given these developments, this study aims to explore how digital mentoring interventions can be adapted to enhance teacher mentorship, improve retention rates, and foster continuous professional growth.

This integrative literature review examines the role of digital mentoring software in teacher development and its effectiveness in enhancing mentorship practices. Specifically, the study aims to: (1) identify key features of mentoring software that contribute to teacher retention, satisfaction, and professional growth; (2) analyze barriers to adopting mentoring software, including financial constraints, institutional resistance, and data privacy concerns; (3) explore lessons from other industries where mentoring software has been successfully implemented; and (4) propose evidence-based recommendations for integrating digital mentoring solutions in educational institutions (Setiani et al., 2024; Bell, 2024).

Despite the potential benefits of mentoring software, its adoption in education is hindered by several challenges. Financial constraints remain a key barrier, limiting many schools' ability to invest in digital mentoring tools (Mumo et al., 2024). Institutional resistance and lack of administrative support further contribute to the failure of mentoring programs to receive necessary resources and prioritization (Tusquellas et al., 2024; Mullen & Hall, 2024). Another challenge is balancing digital technology with human interaction, as effective mentorship relies on interpersonal connections that digital platforms must complement rather than replace (Thornton, 2024). Data privacy and security concerns are also critical, as ensuring compliance with educational data regulations is essential to protect sensitive information (Bagai & Mane, 2024).

This study contributes to the field of Human Resources Development (HRD) and teacher professional development by comprehensively analyzing the role of digital mentoring platforms in enhancing mentorship practices. By incorporating insights from mentorship programs in various sectors, including higher education and professional training, this research seeks to inform educational policymakers, administrators, and software developers on best practices for optimizing digital mentorship in schools (Jayathissa et al., 2024; Krisnaresanti et al., 2024).

## 3. METHODOLOGY

This study employed an integrative literature review approach to evaluate the role of mentoring software in teacher development. The research process began with a comprehensive search of peer-reviewed articles and conference proceedings exclusively using Google Scholar. The search included the keywords "mentoring software in education,digital mentoring platforms,teacher professional development,and mentorship analytics", from 2020 to 2024. The initial search resulted in 961 identified records, as shown in Table 1. These articles were screened for relevance to mentoring software and teacher development, narrowing the selection to 90 records.Full-text articles were further reviewed for alignment with the study's inclusion criteria, reducing the pool to 30 eligible articles. After a final selection process focusing on studies that explicitly addressed mentoring software in education and teacher retention, 17 articles were included in the review, with a focus on the most recent publications.

A thematic analysis was conducted, synthesizing findings from various sectors such as education, healthcare, and corporate industries. This comparative approach helped identify best practices and innovative solutions that could be adapted for educational mentoring software. The synthesis of these studies informed the development of strategic recommendations to improve the adoption and effectiveness of mentoring platforms in educational institutions.

## 4. RELATED WORK

Previous research has explored mentoring software, examining its implementation, benefits, and limitations across industries. Enciso (2024) studied structured mentorship programs in academia, highlighting their role in professional growth and research skills. Similarly, Huang et al. (2024) reviewed technology-enabled teacher development, emphasizing how digital platforms improve mentorship accessibility. Gehreke, Schilling, and Kauffeld (2024) analyzed peer mentoring in higher education, showing how digital tools enhance mentor-mentee relationships and student engagement.

Choudhary et al. (2024) explored technology's role in mentorship, demonstrating how digital tools support career development. Onyema et al. (2024) examined emerging technologies in academic career advancement, focusing on online workshops and data-driven insights. Bell (2024) reviewed e-mentoring for educators, identifying key affordances and constraints.

Bagai and Mane (2024) analyzed AI-powered mentorship platforms, demonstrating how data analytics optimize mentor-mentee pairings. Tusquellas, Palau, and Santiago (2024) examined AI-driven digital mentorship strategies in talent management. Maxwell et al. (2024) synthesized research on mentor education, emphasizing structured mentorship programs.Setiani et al. (2024) reviewed e-mentoring as an alternative to traditional teacher development, showcasing its effectiveness in addressing logistical challenges. Mullen and Hall (2024) explored digital mentorship's role in leadership training. Mumo et al. (2024) studied university-based mentorship, highlighting its impact on career readiness. Zamiri and Esmaeili (2024) identified best practices for integrating digital mentoring into learning communities.Jayathissa et al. (2024) investigated industry mentoring, showing its impact on career success for external degree students. Krisnaresanti et al. (2024) analyzed sustainable teacher development, demonstrating digital mentoring's role in long-term competency building. Anis (2024) examined teacher professional development, addressing educators' evolving needs.

These studies provide a foundation for understanding digital mentorship in education. By synthesizing their findings, this research contributes to the discourse on mentoring software and proposes strategies for enhancing teacher development programs.

## 5. KEY FEATURES OF EDUCATIONAL MENTORING SOFTWARE

Educational mentoring software offers a transformative approach to addressing the challenges of traditional mentorship programs (Enciso, 2024; Huang et al., 2024). By integrating advanced technological features, these platforms streamline the mentoring process, enabling schools to better support teachers despite limited resources (Bagai & Mane, 2024; Choudhary et al., 2024). Features such as mentor-mentee matching, progress tracking, and data-driven insights empower educators to develop professionally while ensuring accountability and engagement (Gehreke, Schilling, & Kauffeld, 2024; Tusquellas, Palau, & Santiago, 2024).

### 5.1. Mentor-Mentee Matching

Effective mentoring relationships begin with compatibility between mentors and mentees. In education, this often involves pairing experienced teachers with new educators. Advanced mentoring software employs algorithms to optimize these pairings by evaluating factors such as teaching styles, subject expertise, and career goals (Choudhary et al., 2024; Onyema et al., 2024). These systems not only improve the quality of matches but also reduce the administrative burden on school administrators, making mentorship programs more accessible and sustainable (Mullen & Hall, 2024).

### 5.2. Progress Tracking and Goal Setting

Mentoring software allows teachers and mentors to set clear objectives and monitor progress through intuitive and user-friendly interfaces. This feature is particularly beneficial for schools, where limited time and competing priorities often hinder consistent and meaningful communication between mentors and mentees. By offering centralized progress tracking dashboards, these platforms ensure that both parties remain aligned on their goals and timelines. Automated reminders further enhance this process by prompting timely check-ins and task completions, fostering a sense of accountability and sustained engagement (Gehreke et al., 2024). For example, digital platforms that integrate analytics provide insights into mentorship quality, participation rates, and program outcomes, empowering school leaders to refine their strategies and identify areas that require targeted interventions (Jayathissa et al., 2024). These metrics contribute to the overall effectiveness and sustainability of mentoring programs by enabling data-driven decision-making (Bagai & Mane, 2024).

### 5.3. Communication Tools

Given the demanding schedules of educators, seamless communication is essential for successful mentoring relationships. Mentoring software integrates features such as instant messaging, video conferencing, and resource sharing to support dynamic interactions between mentors and mentees (Bell, 2024; Setiani et al., 2024). These tools enable educators to connect even in remote or time-constrained settings, ensuring consistent engagement and support. Instant messaging platforms allow for quick exchanges of ideas and resources, bridging gaps that might arise due to physical distance or scheduling conflicts (Mullen & Hall, 2024). Video conferencing, on the other hand, facilitates face-to-face interactions that help maintain the relational aspects of mentoring, which are critical for building trust and rapport (Zamiri& Esmaeili, 2024). Resource-sharing features enable mentors to provide access to curated materials, lesson plans, and professional development content, enhancing the learning experience for mentees (Mumo et al., 2024). Additionally, integrated calendars and scheduling tools streamline the coordination of mentoring sessions, reducing administrative burdens and ensuring that interactions remain consistent and productive (Maxwell et al., 2024).

### 5.4. Analytics and Reporting

Data-driven insights are crucial for evaluating the impact of mentoring programs, as they enable educational institutions to move beyond anecdotal evidence and assess the tangible benefits of their initiatives. Analytics embedded within mentoring software can measure critical outcomes such as teacher satisfaction, retention rates, and professional growth, offering a comprehensive view of a program's effectiveness (Krisnaresanti et al., 2024; Anis, 2024). For example, dashboards and reporting tools allow administrators to identify trends in mentor-mentee interactions, pinpoint areas needing improvement, and recognize successful practices that can be scaled across the institution (Mullen & Hall, 2024). By providing actionable data, these platforms support evidence-based decision-making, enabling schools to allocate resources more efficiently and ensure their mentoring efforts align with broader educational goals (Tusquellas et al., 2024). Additionally, such analytics foster accountability by highlighting participation rates and progress metrics, ensuring that both mentors and mentees remain engaged and committed to achieving their objectives (Jayathissa et al., 2024). These insights not only enhance the quality of mentoring programs but also build a data-driven culture within educational institutions that prioritizes continuous improvement and strategic planning (Setiani et al., 2024).

### 5.5. Benefits of Mentoring Software for Schools

Mentoring software offers numerous benefits for schools, particularly in addressing critical challenges such as teacher retention, professional development, and the creation of collaborative learning environments. These platforms empower educators by providing structured, data-driven approaches that enhance mentoring relationships and ensure accountability (Enciso, 2024; Huang et al., 2024). Mentoring software simplifies administrative tasks for resource-constrained schools, enabling broader program participation despite limited budgets (Mumo et al., 2024; Zamiri& Esmaeili, 2024). By fostering teacher growth and satisfaction, these tools improve individual outcomes and contribute to a more supportive and sustainable educational ecosystem (Jayathissa et al., 2024; Anis, 2024).

### 5.6. Supporting Teacher Development

Mentoring software addresses the critical need for structured professional development in schools. New teachers often face challenges such as classroom management, curriculum planning, and adapting to school culture. Effective mentorship can help bridge these gaps by providing tailored support and helping new educators navigate the complexities of their roles with greater confidence. Digital platforms allow schools to identify areas where teachers need additional training, such as behavior management strategies or integrating technology into the classroom (Gehreke, Schilling, & Kauffeld, 2024). Research on mentoring effectiveness highlights the connection between structured support and improved teacher retention, emphasizing how digital mentorship enhances job satisfaction and efficacy (Tusquellas, Palau, & Santiago, 2024). By aligning mentoring goals with broader educational objectives, schools can ensure that professional development initiatives are both meaningful and impactful (Choudhary et al., 2024).

### 5.7. Enhancing Teacher Retention

Teacher turnover remains a significant issue in education, with financial and emotional costs for schools and students. Mentoring programs supported by software help foster supportive relationships, improving job satisfaction and reducing attrition (Setiani et al., 2024). These programs provide a sense of community and professional belonging that can mitigate the isolation often experienced by new teachers. Research from corporate and healthcare environments demonstrates that structured mentoring programs increase retention rates by providing employees with clear development pathways and ongoing support (Bell, 2024). Schools can adapt these strategies to retain skilled educators by using software to track teacher engagement and identify factors contributing to job dissatisfaction (Maxwell et al., 2024). Studies emphasize that fostering a supportive mentoring culture not only benefits individual teachers but also strengthens the overall school environment (Mullen & Hall, 2024).

### 5.8. Building Collaborative Communities

Mentoring software fosters a culture of collaboration by connecting teachers across departments, schools, or even districts. These connections break down silos and encourage the sharing of best practices, lesson plans, and professional insights. Digital platforms facilitate the creation of virtual communities where educators can engage in peer mentoring, share innovative teaching strategies, and collaborate on cross-disciplinary projects (Mullen & Hall, 2024; Krisnaresanti et al., 2024). Such networks provide teachers with a support system that extends beyond their immediate school environment, fostering professional growth and a stronger sense of belonging. Collaborative mentoring environments also encourage experienced teachers to take on leadership roles, further enriching the professional development landscape within schools (Anis, 2024). These communities not only benefit individual teachers but also contribute to a more cohesive and innovative educational culture, ultimately improving outcomes for both educators and students (Mumo et al., 2024).

## 6. Challenges in Adopting Mentoring Software in Education

While mentoring software has transformative potential for schools, its adoption is not without significant challenges. Many educational institutions struggle with financial barriers, as high licensing fees and implementation costs can be prohibitive for resource-constrained districts (Mumo et al., 2024; Zamiri& Esmaeili, 2024). A lack of institutional support often hinders the effective integration of these tools, as school leaders may undervalue mentoring programs or fail to prioritize necessary training for staff (Mullen & Hall, 2024; Maxwell et al., 2024). Balancing technological efficiency with the relational aspects of mentoring poses a challenge, as an over-reliance on software may detract from the human connections essential to mentorship (Setiani et al., 2024; Bell, 2024). Concerns over data privacy and compliance with regulations such as FERPA present additional hurdles for schools seeking to adopt these platforms (Bagai & Mane, 2024; Tusquellas, Palau, & Santiago, 2024).

### 6.1. Financial Constraints

Many schools operate with limited budgets, making it difficult to invest in mentoring software including licensing fees, training costs, and ongoing maintenance can be prohibitive(Mumo et al., 2024). Zamiri and Esmaeili (2024) suggest that cost-effective ICT tools and open-source platforms may provide viable alternatives for resource-constrained schools. Additionally, partnerships with educational nonprofits or leveraging government grants can help mitigate these financial barriers. Schools that successfully integrate mentoring software often do so by prioritizing scalable solutions and incremental implementation, ensuring affordability while still reaping the benefits of digital mentoring systems.

### 6.2. Balancing Technology and Human Interaction

While mentoring software offers numerous advantages, it cannot fully replace the human connection that underpins effective mentoring relationships. Over-reliance on digital tools may detract from the personal aspects of mentorship, such as emotional support and trust-building (Setiani et al., 2024). Effective mentoring relies on nuanced interpersonal interactions that technology cannot replicate entirely. Bell (2024) highlights the importance of integrating digital tools as a complement to, rather than a replacement for, traditional mentoring practices. Schools must balance leveraging technology and preserving the relational elements of mentoring by encouraging in-person meetings, fostering emotional connections, and using digital tools to enhance rather than substitute face-to-face interactions. Combining these approaches can create a hybrid mentoring model that maximizes the strengths of both traditional and digital methodologies.

### 6.3. Lessons from Other Industries

The adoption of mentoring software in other industries provides valuable lessons for the education sector. In healthcare, mentoring platforms have been instrumental in onboarding new professionals, reducing turnover, and improving patient outcomes, demonstrating how structured programs can address sector-specific challenges (Choudhary et al., 2024; Onyema et al., 2024). Similarly, corporate environments leverage mentoring analytics to enhance employee engagement and leadership development, showcasing the role of data-driven insights in achieving organizational goals (Mullen & Hall, 2024; Tusquellas, Palau, & Santiago, 2024). By examining these successful applications, schools can identify strategies for implementing mentoring software that balances efficiency, personalization, and scalability.

### 6.4. Healthcare

In the healthcare industry, mentoring programs supported by software have been used to onboard new employees and improve patient care outcomes. Digital mentoring platforms enable hospitals to match mentors with mentees based on specialties and career goals, providing targeted support (Choudhary et al., 2024). These programs address the high-stress environment of healthcare by ensuring that new professionals have access to experienced mentors who can guide them through complex processes and emotional challenges. The data-driven features of these platforms also allow healthcare administrators to monitor program effectiveness, identify areas of improvement, and tailor mentoring strategies to the unique needs of their staff (Onyema et al., 2024). Schools can draw parallels by using similar systems to support new teachers in their specific subject areas, offering targeted guidance that enhances both professional confidence and instructional quality.

### 6.5. Corporate Sector

Many corporations use mentoring software to enhance employee engagement and retention. For example, Mullen and Hall (2024) describe how analytics-driven mentoring programs help organizations identify high-potential employees and provide tailored development plans. Such programs often incorporate advanced reporting features that track employee progress, highlight skill gaps, and offer insights into team dynamics. By adapting these strategies, schools can identify and support teachers who demonstrate leadership potential, fostering career advancement and reducing burnout. Additionally, corporate mentoring models often emphasize cross-functional relationships, where employees from different departments collaborate and learn from each other. Educational institutions can replicate this by fostering interdisciplinary mentoring among teachers, encouraging collaboration across grade levels and subject areas to enrich professional development.

## 7. Future Scope and Improvements

Although mentoring software has shown considerable potential in enhancing teacher development, further improvements are necessary. One promising advancement is the integration of artificial intelligence (AI) and machine learning to personalize mentor-mentee matching and predict mentorship outcomes (Bagai & Mane, 2024). AI-driven analytics will likely provide deeper insights into mentoring effectiveness and recommend individualized development opportunities

Data privacy and security remain pressing concerns, particularly in educational settings where compliance with regulations such as FERPA is essential. Future developments couldstrengthen encryption protocols, decentralize data storage, and enhance user-controlled privacy settings to protect sensitive information (Setiani & Chaeruman, 2024).

Future research should focus on longitudinal studies to assess the long-term impact of mentoring software on teacher retention, job satisfaction, and student outcomes. These studies will help refine best practices and provide valuable insights for policymakers and educational leaders (Jayathissa et al., 2024).The next generation of mentoring software can improve scalability, effectiveness, and inclusivity by addressing these areas, ultimately fostering a more equitable and efficient education system.

## 8. LIMITATIONS AND CHALLENGES

Despite its advantages, mentoring software faces several limitations and challenges. One significant barrier is the high cost associated with implementation and maintenance, particularly for schools with constrained budgets (Enciso, 2024). Many institutions struggle to afford licensing fees, ongoing technical support, and staff training, limiting their ability to adopt and sustain mentoring programs (Mumo et al., 2024).

Another challenge is the potential for reduced interpersonal connections in digital mentoring environments. While virtual tools facilitate communication, they may not fully replicate the relational aspects of traditional face-to-face mentoring (Setiani&Chaeruman, 2024). Effective mentorship often relies on trust and personal interactions, which digital platforms must complement rather than replace (Bell, 2024). Schools must balance the use of technology with strategies that preserve the human aspects of mentorship, such as encouraging occasional in-person meetings and fostering strong mentor-mentee relationships (Maxwell et al., 2024).

Addressing these limitations will require ongoing improvements in mentoring software design, affordability, and security to ensure broader adoption and long-term success in educational settings. By strategically overcoming these challenges, schools can unlock the full potential of digital mentoring to enhance teacher development and professional growth.

## 9. CONCLUSION

Mentoring software represents a transformative opportunity for supporting teacher development and addressing the challenges faced by schools. These platforms provide structured, data-driven solutions that enhance traditional mentoring methods by offering goal setting, progress tracking, and advanced analytics. The integration of these features ensures accountability and fosters meaningful mentor-mentee relationships, ultimately driving teacher satisfaction, professional growth, and retention.

While mentoring software presents significant benefits, its widespread adoption is hindered by financial constraints, institutional resistance, and data privacy concerns. However, case studies from other industries, such as healthcare and corporate environments, demonstrate that strategic implementation, flexible funding models, and strong leadership commitment can overcome these barriers. Schools can adopt best practices from these sectors to develop sustainable and effective mentoring programs that maximize the advantages of digital mentorship.

Table 1: PRISMA Table for Mentoring Software Literature Review

| Stage | Description | Outcome |
|---|---|---|
| **Identification** | Database search of peer-reviewed articles on mentoring software, with a focus on education. | 906 records identified |
| **Screening** | Abstracts and titles screened for relevance to mentoring software and teacher development. | 90 records screened |
| **Eligibility** | Full-text articles reviewed for alignment with inclusion criteria (e.g., focus on education). | 30 articles eligible |
| **Included** | Final selection of articles with a focus on mentoring software in education and teacher retention. | 17 articles included |